\documentclass[PRM,twocolumn,groupedaddress,nofootinbib,superscriptaddress,longbibliography]{revtex4-2}
\usepackage[version=3]{mhchem}
\usepackage{hyperref}
\usepackage{graphicx}
\usepackage{amsmath,amsfonts}
\usepackage{dcolumn}
\usepackage{bm}
\usepackage{color}
\usepackage{multirow}
\usepackage{physics}
\usepackage{subfigure}
\usepackage[utf8]{inputenc}
\usepackage{titlesec}
\usepackage{float}
\usepackage[margin=0.6in]{geometry}
\usepackage[dvipsnames]{xcolor}

\begin{document}


\clearpage

\title{Anisotropic magnetism and Kondo-lattice behavior in the frustrated antiferromagnet \ce{Ce3MgBi5}} 

\author{Karolina Gornicka}
\email{gornickaka@ornl.gov}
\affiliation{Materials Science and Technology Division, Oak Ridge National Laboratory, Oak Ridge, Tennessee 37831, United States}
\affiliation{Faculty of Applied Physics and Mathematics and Advanced Materials Centre, Gdansk University of Technology, ul. Narutowicza 11/12, 80-233 Gdańsk, Poland}

\author{Brenden R. Ortiz}
\affiliation{Materials Science and Technology Division, Oak Ridge National Laboratory, Oak Ridge, Tennessee 37831, United States}

\author{Matthew S. Cook}
\affiliation{Materials Science and Technology Division, Oak Ridge National Laboratory, Oak Ridge, Tennessee 37831, United States}

\author{Heda Zhang}
\affiliation{Materials Science and Technology Division, Oak Ridge National Laboratory, Oak Ridge, Tennessee 37831, United States}

\author{Andrew D. Christianson}
\affiliation{Materials Science and Technology Division, Oak Ridge National Laboratory, Oak Ridge, Tennessee 37831, United States}

\author{Andrew F. May}
\affiliation{Materials Science and Technology Division, Oak Ridge National Laboratory, Oak Ridge, Tennessee 37831, United States}

\date{\today}

\begin{abstract}
We report the synthesis and physical characterization of single-crystalline \ce{Ce3MgBi5}, a previously unexplored member of the \ce{Ce3\textit{M}\textit{Pn}5} family. This compound crystallizes in the hexagonal \textit{P}6$_3$/\textit{mcm} structure, featuring an anisotropic Ce sublattice composed of zig-zag chains along the \textit{c} axis and a distorted kagome-like network in the basal plane. Magnetization measurements reveal antiferromagnetic order below $T_N \approx 4.2~K$ with strong magnetic anisotropy and multiple field-induced metamagnetic transitions for fields applied perpendicular to $[001]$, leading to a dome-shaped $H–T$ phase diagram. Electrical transport exhibits characteristic signatures of a Ce-based Kondo lattice, including broad resistivity maxima and pronounced field-dependent anomalies in the magnetoresistance and Hall response that track the magnetic phase boundaries. Specific-heat measurements confirm the magnetic transition and show that the full $R\ln2$ entropy expected for a Ce$^{3+}$ Kramers doublet is recovered by 20 K, indicating an extended temperature range of magnetic fluctuations consistent with Kondo correlations. Our results establish \ce{Ce3MgBi5} as a platform within the \ce{Ce3\textit{M}\textit{Pn}5} family for exploring the interplay of geometric frustration, magnetic anisotropy, and Kondo-lattice physics under applied magnetic fields.
\end{abstract}


\maketitle

\section{Introduction}
 Intermetallic compounds containing rare-earth elements often exhibit complex magnetic behavior due to interactions between localized $4f$ moments and conduction electrons \cite{van1962note, kasuya1956theory, ruderman1954indirect}, with Ce-based compounds providing a particularly rich playground owing to the presence of a single unpaired $4f$ electron and valence fluctuations. In such systems, the magnetic ground state is commonly discussed in terms of the interplay between the Ruderman–Kittel–Kasuya–Yosida (RKKY) interaction, which tends to stabilize long-range magnetic order, and the Kondo effect, which screens local moments to form a nonmagnetic Fermi liquid \cite{doniach1977kondo}. The balance between these interactions is strongly influenced by crystal electric field (CEF) effects, which introduce magnetic anisotropy and modify the effective moment of the Ce ions. As a result, many Ce systems exhibit long-range magnetic order at low temperatures together with transport and thermodynamic signatures that deviate from simple localized-moment behavior \cite{yang2008scaling, gupta1983magnetic, mallik1997kondo, zhang2024structural, ghimire2016physical, shishido2010tuning, macaluso2003synthesis, ye2022magnetic, zhang2020structural, he2025rich, hundley1994substitutional, motoyama2018magnetic}. 

A prominent materials platform for exploring the interplay of RKKY interactions, CEF effects, and Kondo hybridization is the \ce{Ce3\textit{M}\textit{Pn}5} family (\textit{M} = transition metal or Mg/Sc, \textit{Pn} = Sb, Bi, or As), which crystallizes in a hexagonal structure with space group \textit{P}6$_3$/\textit{mcm} \cite{khoury2024towards,khoury2022ln3mbi5,zhang2024crystal,murakami2017hypervalent,moore2002physical,han2023complex,wang2022one,xu2024magnetization,gauthier2024magnetic,nakagawa2023single,zelinska2008ternary,ovchinnikov2018undistorted, fu2024high, min2025synthesis, ritter2021magnetic}. As shown in Fig.\ref{Fig.1}(a), the crystal structure comprises two quasi-one-dimensional (quasi-1D) substructures aligned along the crystallographic \textit{c} axis - face-sharing \ce{\textit{MPn}6} octahedra and hypervalent \textit{Pn} chains - which play an important role in shaping the electronic properties. In particular, the hypervalent bonding within the \textit{Pn} chains introduces additional electronic complexity and contributes to the anisotropic character of the transport response \cite{khoury2024towards, khoury2022class}. Within this structural framework, however, the magnetic behavior is primarily governed by the arrangement of the Ce ions. Along the \textit{c} axis, the Ce atoms form zig-zag chains with relatively short nearest-neighbor distances, suggesting enhanced magnetic exchange interactions along this direction. In contrast, within the \textit{ab} plane the Ce atoms form a distorted kagome-like network, which can give rise to competing magnetic interactions due to the lattice geometry. Notably, while the RKKY interaction provides the dominant mechanism for coupling the localized Ce $4f$ moments via conduction electrons, its effective strength and sign depend intimately on both interatomic distances and details of the electronic structure. As a result, RKKY interactions do not simply scale with nearest-neighbor separations and can remain significant over longer distances, allowing both interplane and in-plane exchange pathways to contribute to the magnetic ground state.

Recent studies of the \ce{Ce3\textit{M}\textit{Pn}5} family have shown that several members exhibit antiferromagnetic (AFM) ground states with pronounced magnetic anisotropy, while Kondo hybridization can significantly influence the magnetic and transport properties without fully suppressing long-range order \cite{motoyama2018magnetic,he2025rich, xu2024magnetization, matin2017probing, fu2024high, hayami2022magnetic,park2026magneticorderexcitationsce3tibi5}. A representative example is the Kondo Weyl semimetal candidate \ce{Ce3TiSb5}, where the quasi-kagome arrangement of Ce moments leads to strong geometric frustration and a complex magnetic phase diagram \cite{matin2017probing}. Recently, X. He \textit{et al}. \cite{he2025rich} reported that \ce{Ce3TiSb5} exhibits multiple field- and temperature-driven magnetic transitions together with unconventional transport responses, underscoring the potential of the \ce{Ce3\textit{M}\textit{Pn}5} family to host nontrivial correlated-electron behavior. Similar phenomena have also been reported in other members of the family, including \ce{Ce3ScBi5} \cite{xu2024magnetization} and \ce{Ce3TiBi5} \cite{motoyama2018magnetic, hayami2022magnetic, gauthier2024magnetic}, which exhibit anisotropic antiferromagnetic order and rich field–temperature phase diagrams, reflecting the combined influence of lattice geometry, crystal electric field anisotropy, and Kondo hybridization on the magnetic ground state.

Here, we report the synthesis of high-quality single crystals of \ce{Ce3MgBi5}, a previously unreported member of the \ce{Ce3\textit{M}\textit{Pn}5} family, and present a comprehensive investigation of its structural, magnetic, thermodynamic, and transport properties. Our results reveal antiferromagnetic order below $T_N \approx 4.2~\mathrm{K}$, accompanied by pronounced magnetic anisotropy and multiple field-induced metamagnetic transitions for magnetic fields applied perpendicular to the $[001]$ direction, indicating a delicate balance between multiple energy scales. Electrical transport measurements exhibit strong magnetic-field dependence and low-temperature anomalies that correlate with the magnetic phase behavior. In addition, the magnetic entropy released at the ordering transition is reduced compared to the value expected for a fully localized Ce$^{3+}$ doublet, as commonly observed in systems where Kondo screening of the Ce moments is relevant. Taken together, these findings establish \ce{Ce3MgBi5} as a platform for exploring the interplay between geometric frustration, Kondo physics, and anisotropic magnetism within the \ce{Ce3\textit{M}\textit{Pn}5} family. 

\section{Experimental details}
\begin{figure}[t]
\centering
\includegraphics[width=1\columnwidth]{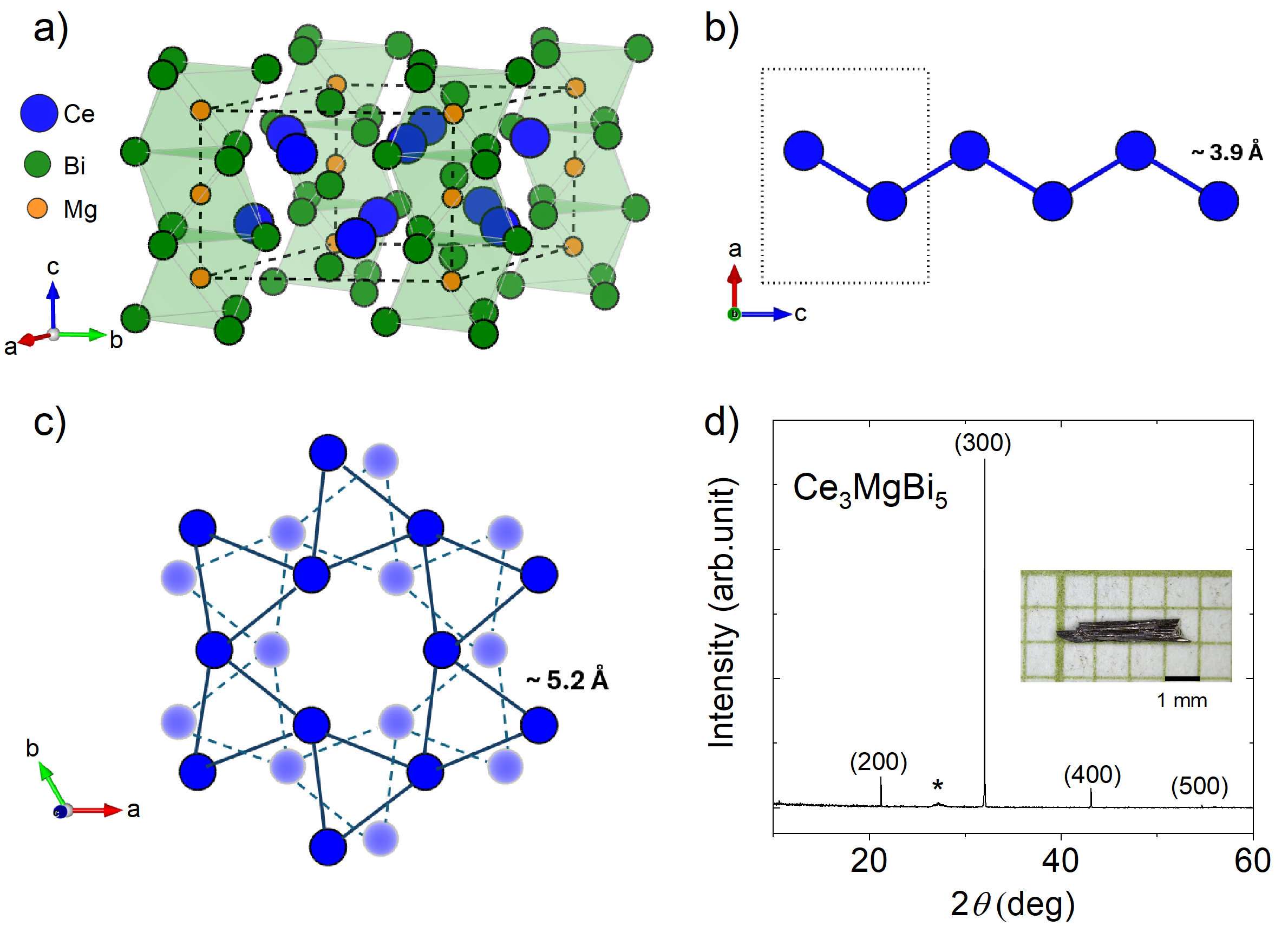}
\caption{(a) Crystal structure of \ce{Ce3MgBi5} highlighting the arrangement of Ce (blue), Bi (green), and Mg (orange) atoms. The structure consists of face-sharing \ce{MgBi6} octahedra and hypervalent Bi chains running along the \textit{c} axis. (b) Zig-zag chain of Ce atoms along the \textit{c} direction with a nearest-neighbor Ce–Ce distance of approximately 3.9 ~\AA. (c) Arrangement of Ce atoms within the \textit{ab} plane, showing two adjacent Ce layers that together form a distorted kagome-like network, with in-plane Ce–Ce bonds of about 5.2 ~\AA. (d) X-ray diffraction pattern collected from the as-grown facets of a \ce{Ce3MgBi5} single crystal. The peak marked with an asterisk (*) originates from residual Bi flux on the crystal surface. The inset shows a photograph of \ce{Ce3MgBi5} single crystal with [001] oriented along the long-axis of the columnar morphology.}
\label{Fig.1}
\end{figure}

Single crystals of \ce{Ce3MgBi5} were grown using a Mg–Bi self-flux method. High-purity elemental Ce (99.9\%), Mg (99.9\%), and Bi (99.999\%) were weighed in an initial molar ratio of Ce:Mg:Bi = 1:6:9 and loaded into 2 mL Canfield crucibles fitted with a catch crucible and a porous frit \cite{canfield2016use}. The crucible set  was then sealed in a quartz tube under vacuum. The sealed ampule was placed in a box furnace, heated to 900$^\circ$C, and held at this temperature for 24 h to ensure complete melting and homogenization of the melt. The temperature was slowly cooled to 700$^\circ$C at a rate of 2$^\circ$C/h, after which the ampoule was removed from the furnace and the excess flux was separated by centrifugation. As shown in the inset of Fig.\ref{Fig.1}(d), rodlike single crystals with typical lengths of 2–5 mm and masses ranging from approximately 10 to 60 mg were obtained. The crystals are moderately stable under ambient conditions, although prolonged exposure to air leads to noticeable surface degradation. Single crystals of \ce{La3MgBi5}, used as a nonmagnetic analog, were grown using an analogous method.

The chemical homogeneity and elemental composition of the single crystals were examined by energy-dispersive x-ray spectroscopy (EDS) using a Hitachi TM3000 scanning electron microscope equipped with a Bruker Quantax 70 EDS system. The analysis confirmed the expected 3:1:5 stoichiometry (Ce:Mg:Bi = 34 $\pm$ 1 : 11 $\pm$ 2 : 55 $\pm$ 2) within the experimental uncertainty and revealed no detectable secondary phases.

Single-crystal x-ray diffraction measurements were performed on small, regularly shaped crystals mounted on Kapton loops using glycerol as an adhesive. Diffraction data were collected on a Bruker D8 Advance Quest diffractometer equipped with a graphite monochromator and using Mo K$\alpha$ radiation ($\lambda = 0.71073$ Å). The crystal structure was solved by direct methods and refined using full-matrix least-squares procedures implemented in the Bruker APEX4 software package. To verify the crystallographic orientation of the as-grown crystals, additional x-ray diffraction measurements were carried out on flat crystal facets using a PANalytical X’Pert Pro diffractometer with monochromated Cu K$\alpha_1$ radiation. Due to the extreme air sensitivity of the powdered samples, which decomposed almost immediately upon exposure to air, powder x-ray diffraction measurements were not performed.

Magnetization measurements were carried out using a Quantum Design Magnetic Property Measurement System (MPMS3) based on a superconducting quantum interference device (SQUID), operated in vibrating-sample magnetometry (VSM) mode. Measurements were carried out over the temperature range $1.9~\mathrm{K} \le T \le 300~\mathrm{K}$ in magnetic fields up to $H = 70~\mathrm{kOe}$. The external magnetic field was applied either parallel to the normal vector of the as-grown facet, labeled as $H \perp [001]$, or parallel to the \textit{c} axis ($H \parallel [001]$). Low-field magnetization as a function of temperature, $M(T)$, was measured under zero-field-cooled (ZFC) warming and field-cooled (FC) cooling protocols. Additional high-field magnetization measurements up to $12~\mathrm{T}$ were performed using a Quantum Design DynaCool Physical Property Measurement System (PPMS).

\begin{figure*}[ht!]
\centering
\includegraphics[width=1\textwidth]{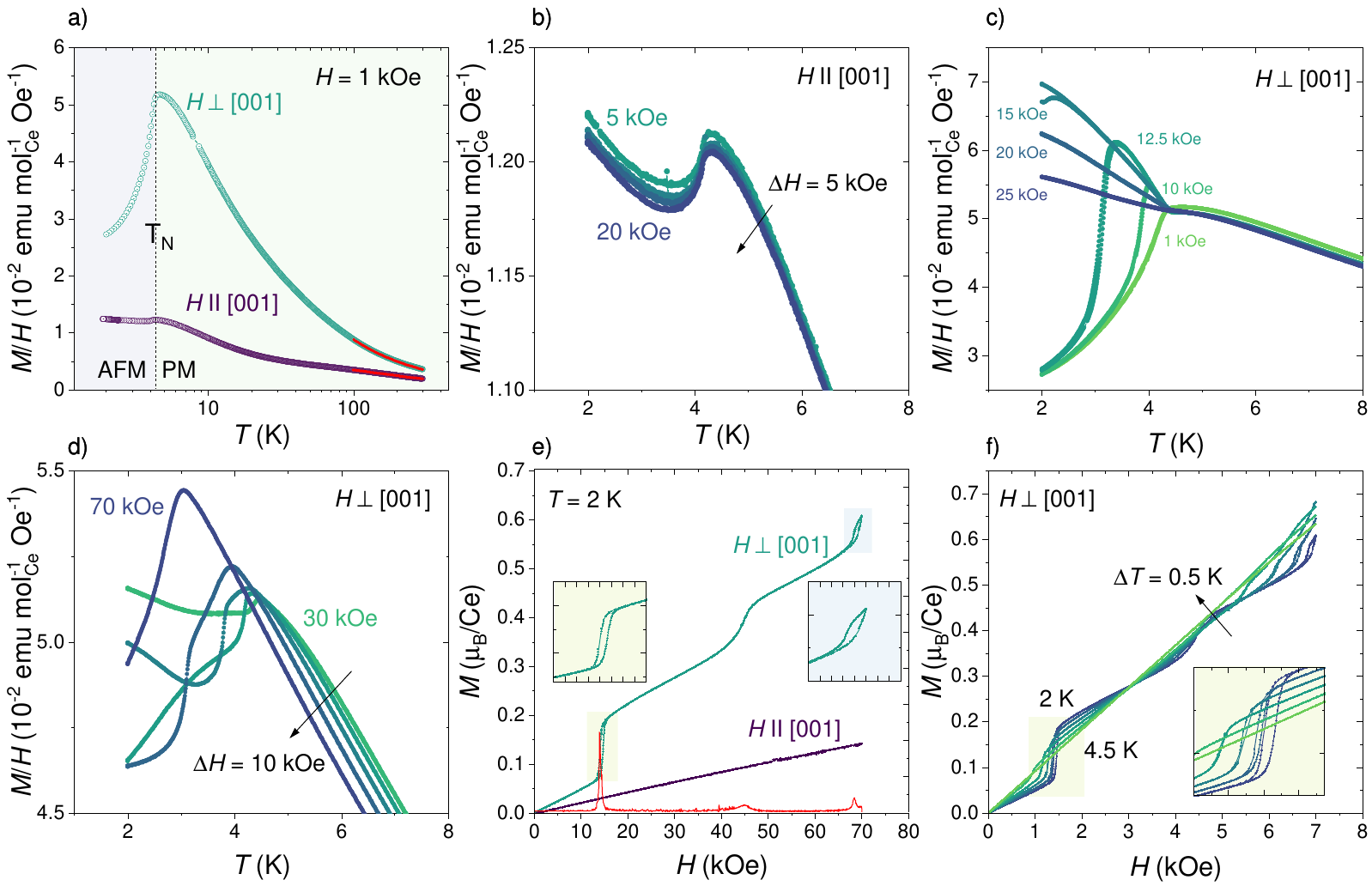}
\caption{(a) The magnetic susceptibility, $\textit{M}/\textit{H}$, as a function of temperature measured in an applied field of 1 kOe for $H \perp [001]$ and $H \parallel [001]$, together with the Curie-Weiss fits (red lines). The vertical dashed line marks the antiferromagnetic transition temperature $T_N$. (b) FC magnetic susceptibility data measured for $H \parallel [001]$ under magnetic fields between 5 and 20 kOe ($\Delta\textit{H} = 5~\mathrm{kOe}$). (c) $\textit{M}/\textit{H}$ vs $\textit{T}$ for $H \perp [001]$ under magnetic fields ranging from 1 kOe to 25 kOe, presenting the evolution of the magnetic transition with increasing field. (d) Low-temperature magnetization data for $H \perp [001]$ at higher fields ($\Delta\textit{H} = 10~\mathrm{kOe}$). (e)  Field-dependent magnetization at 2 K for \ce{Ce3MgBi5}. The derivative, $\textit{dM}/\textit{dH}$ (red line, arbitrary units), emphasizes three metamagnetic transitions observed for $H \perp [001]$. Insets show enlarged views of first and third metamagnetic transitions. (f) $\textit{M}(\textit{H}$) curves for $H \perp [001]$ at selected temperatures between 2 and 4.5 K ($\Delta\textit{T} = 0.5~\mathrm{K}$).} 
\label{fig:2}
\end{figure*}

Electrical resistivity ($\rho$) and magnetoresistance (MR) measurements were carried out in the PPMS using a standard four-probe configuration. Platinum wires were attached to the samples using silver paste (DuPont~4929N). The electrical current was applied along the crystallographic \textit{c} direction, while the magnetic field was applied perpendicular to the \textit{c} axis. Hall effect measurements were conducted using the ac resistance (ETO) option of the DynaCool PPMS. The raw transverse resistance data were antisymmetrized with respect to the magnetic field to remove contributions from longitudinal magnetoresistance; no additional background subtraction or corrections were applied. Heat-capacity ($C_p$) measurements were performed using the relaxation method in the PPMS. Field-dependent measurements were carried out for magnetic fields applied perpendicular to the $[001]$ direction. Due to experimental constraints related to sample stability on the calorimetry platform under high magnetic fields, the applied field in the heat-capacity measurements was limited to $H \leq 30~\mathrm{kOe}$. Additional low-temperature heat-capacity measurements down to the $^{3}$He temperature range were carried out in zero magnetic field.

\section{Results and discussion}

Single-crystal x-ray diffraction reveals that \ce{Ce3MgBi5} crystallizes in the hexagonal \textit{P}6$_3$/\textit{mcm} structure, consistent with other members of the \ce{\textit{Ln}3MgBi5} (\textit{Ln} = lanthanide elements) family. The refined lattice parameters are $a = b = 9.67(2)$~\AA\ and $c = 6.477(2)$~\AA, which lie between those reported for \ce{La3MgBi5} \cite{pan2006synthesis} and \ce{Pr3MgBi5} \cite{han2023complex}, following the expected lanthanide-contraction trend. The resulting CIF file is included in the Supplemental Material \cite{Supplemental}. In \ce{Ce3MgBi5}, the arrangement of the Ce ions is strongly anisotropic. The nearest-neighbor Ce–Ce distance along the crystallographic \textit{c} axis is approximately $3.9$~\AA\ (Fig.\ref{Fig.1}(b)), which is significantly shorter than the corresponding distance of about $5.2$~\AA\ within the \textit{ab} plane (Fig.\ref{Fig.1}(c)). Notably, while the crystal structure is globally centrosymmetric, local inversion symmetry is broken at the Ce sites, a feature that permits additional anisotropic exchange terms and may have important consequences for the magnetic interactions discussed below \cite{motoyama2018magnetic, hayami2022magnetic,park2026magneticorderexcitationsce3tibi5}.

X-ray diffraction collected from the flat facet of the as-grown single crystal reveals only sharp $(h00)$ reflections (see Fig.\ref{Fig.1}(d)), indicating that the crystals grow with the crystallographic $[001]$ direction aligned along the long axis. A weak additional reflection, marked by an asterisk, is attributed to residual Bi flux remaining on the crystal surface.

In order to reveal the magnetic response of \ce{Ce3MgBi5}, the magnetization, $\textit{M}/\textit{H}$, was measured in an applied field of $\textit{H} = 1~\mathrm{kOe}$ for both directions over the temperature range $1.9~\mathrm{K} \le T \le 300~\mathrm{K}$, as shown in Fig.\ref{fig:2}(a). The data are plotted on a logarithmic temperature scale to better visualize both the high- and low-temperature regimes. At low temperatures, both magnetization curves exhibit a clear anomaly associated with the onset of long-range antiferromagnetic order. The Néel temperature was determined more precisely from the maximum in $\textit{d}(\textit{MT})/\textit{dT}$, giving $T_N = 4.23~\mathrm{K}$. For $H \perp [001]$, the magnetization shows a pronounced decrease upon cooling below the ordering temperature, whereas the $\textit{M}(\textit{T})$ for $H \parallel [001]$ exhibits only a weak temperature dependence across the transition. The suppression of $M(T)$ for $H \perp [001]$ is consistent with the onset of antiferromagnetic correlations involving in-plane components of the Ce moments, while the nearly unchanged response along $[001]$ reflects the hard-axis character of the \textit{c} direction.

At high temperatures, the magnetic susceptibility for both field orientations follows Curie–Weiss behavior, and the data above 100 K were fitted using a modified Curie–Weiss expression ($\chi(\textit{T}) = \chi_0 + \textit{C}/(\textit{T} - \Theta_p)$, where $\textit{C}$ is the Curie constant). For $H \perp [001]$, the fit yields an effective magnetic moment $\mu_\text{eff} = 2.72(1)~\mu_B/Ce$ and a Weiss temperature $\Theta_p = -12.8(1)~\mathrm{K}$, while for $H \parallel [001]$ we obtained $\mu_\text{eff} = 2.76(1)~\mu_B/Ce$ and a much larger negative Weiss temperature $\Theta_p = -165.1(6)~\mathrm{K}$. The extracted effective moments are close to the theoretical value expected for a free Ce$^{3+}$ ion ($2.54~\mu_B$), confirming the predominantly localized nature of the Ce $4\textit{f}$ moments. The negative values of $\Theta_p$ point to dominant antiferromagnetic interactions between the localized Ce moments. However, the fact that long-range antiferromagnetic order develops at temperatures far below $|\Theta_p|\sim165~\mathrm{K}$ (along $[001]$) suggests that $\Theta_p$ is strongly influenced by crystal electric field effects and possibly Kondo interactions. The large difference between the Weiss temperatures obtained for the two field orientations reflects strong magnetic anisotropy, which we tentatively attribute primarily to CEF effects. This attribution is supported by recent neutron scattering work on the related materials \ce{Ce3TiBi5} and \ce{Ce3ZrBi5} which identifies crystal field excitations at approximately 12 and 27 meV in both materials \cite{park2026magneticorderexcitationsce3tibi5}. Similar anisotropic Curie–Weiss behavior has been reported in other members of the \ce{Ce3\textit{MPn}5} family, including \ce{Ce3TiSb5} \cite{matin2017probing}, \ce{Ce3ScBi5} \cite{xu2024magnetization}, \ce{Ce3TiBi5} \cite{motoyama2018magnetic}, and \ce{Ce3VAs5} \cite{min2025synthesis}. 

To further investigate the magnetic transition and its evolution with applied magnetic field, the temperature dependence of the magnetization measured under different fields for $H \parallel [001]$ is shown in Fig.\ref{fig:2}(b). For clarity, only data up to $20~\mathrm{kOe}$ are presented. No measurable difference was observed between zero-field-cooled and field-cooled modes for this orientation; therefore, only FC data are shown. A clear cusp associated with the antiferromagnetic transition is observed in $\textit{M}(\textit{T})$, with the corresponding transition temperature $T_N$ remaining nearly field independent. Upon further cooling below $T_N$, the magnetization shows a slight increase, in contrast to the response for $H \perp [001]$. Similar low-temperature behavior has been reported for the isostructural compound \ce{Nd3ScBi5} \cite{gornicka2025anisotropic}, which has been attributed to crystal electric field effects affecting the magnetic response along the hard axis.

Figures~\ref{fig:2}(c) and (d) summarize the temperature dependence of the magnetization measured for $H \perp [001]$ under ZFC and FC conditions over different magnetic-field ranges. For clarity, Fig.\ref{fig:2}(c) shows selected data measured in fields up to 25 kOe, while Fig.\ref{fig:2}(d) focuses on higher fields between 30 and 70 kOe. 

At low applied fields ($H< 15~\mathrm{kOe}$), the antiferromagnetic transition is clearly visible as a cusp in $\textit{M}(\textit{T})$, which shifts to lower temperatures with increasing field, consistent with the progressive suppression of antiferromagnetic order. At the same time, the magnitude of the cusp gradually increases, reflecting the enhanced magnetic polarization of the ordered state. In this low-field regime, the ZFC and FC magnetization curves largely overlap, with a noticeable separation appearing only at $H = 15~\mathrm{kOe}$, indicative of field-induced changes in the magnetic state. As the magnetic field is increased further ($15-25~\mathrm{kOe})$, the anomaly associated with $T_N$ becomes significantly weakened, and the magnetization exhibits a slight increase upon cooling through the transition. This behavior may reflect a field-induced modification of the magnetic structure, although the microscopic details cannot be determined without more advanced experimental studies.  

\begin{figure*}[ht!]
\centering
\includegraphics[width=1\textwidth]{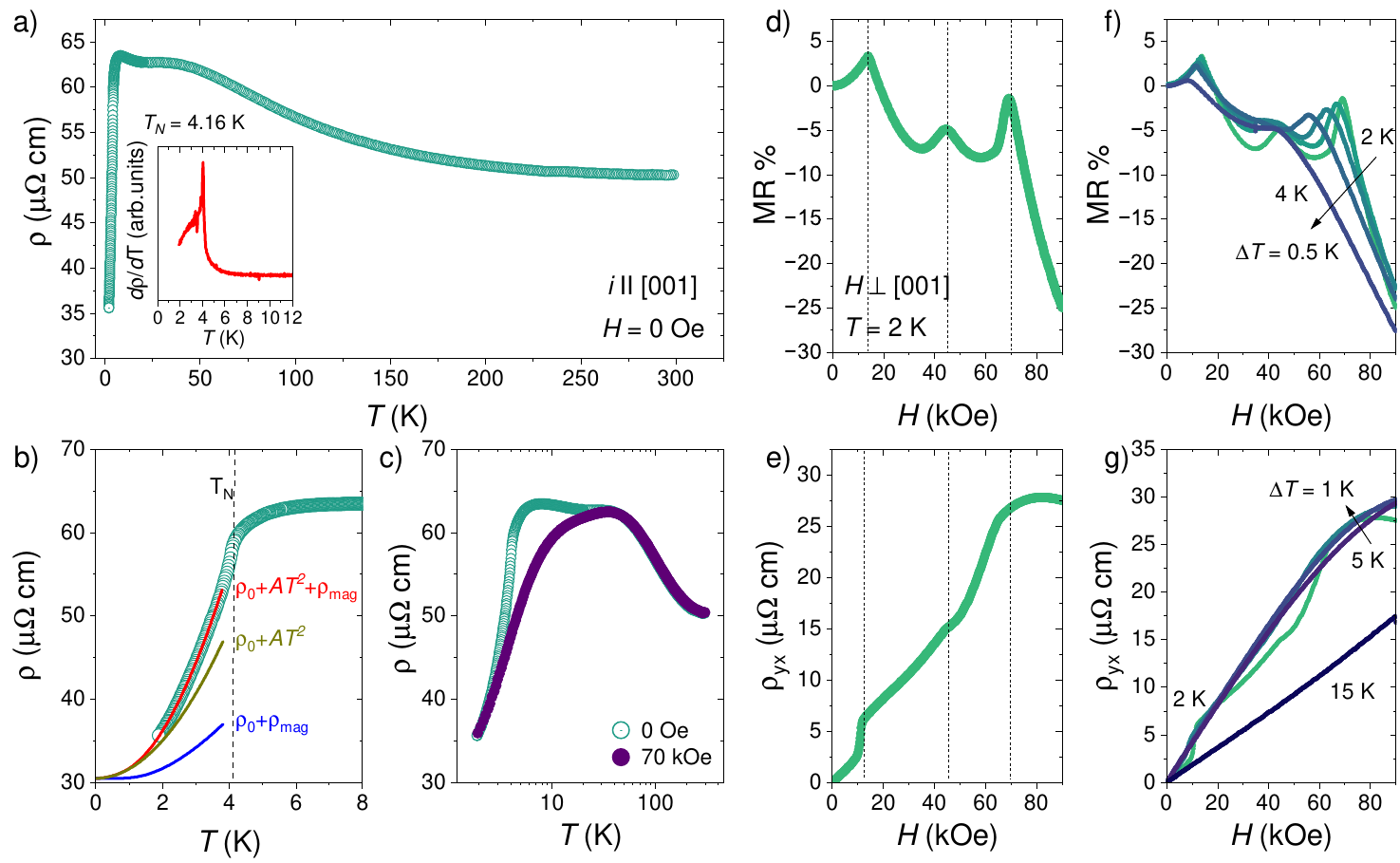}
\caption{(a) Temperature dependence of the electrical resistivity $\rho(T)$ of \ce{Ce3MgBi5} measured in zero magnetic field from $300~$ to $1.9~\mathrm{K}$. The inset shows the temperature derivative $\mathrm{d}\rho/\mathrm{d}T$, highlighting the antiferromagnetic transition at $T_N = 4.16~\mathrm{K}$. (b) Enlarged view of the low-temperature region of $\rho(T)$ in zero field. The red solid line represents a fit to the antiferromagnetic spin-wave scattering model, with $\Delta$ fixed to the value obtained from heat-capacity data. The auxiliary curves highlight the relative contributions of the $\rho_0 + AT^2$ and magnon terms. The portions of the curves below 1.8 K are extrapolations (see text). (c) $\rho(T)$ measured at $H = 0$ and $70~\mathrm{kOe}$, plotted on a logarithmic temperature scale to emphasize the suppression of the low-temperature resistivity maximum under applied magnetic field. (d) Magnetoresistance MR = [$\rho(H)-\rho(0)$]/$\rho(0)$ measured at $T=2~\mathrm{K}$ for $H \perp [001]$, showing pronounced anomalies associated with metamagnetic transitions. (e) Hall resistivity $\rho_{yx}(H)$ at $T=2~\mathrm{K}$. (f) Temperature evolution of the MR in the vicinity of the magnetic ordering region ($\Delta T = 0.5~\mathrm{K}$). (g) The Hall resistivity for selected temperatures ($\Delta T = 1$ K), illustrating the gradual suppression of nonlinear features with increasing temperature.} 
\label{fig:3}
\end{figure*}

A further increase of the magnetic field restores a decrease of $\textit{M}(\textit{T})$ upon cooling below the transition, while the cusp associated with $T_N$ continues to shift toward lower temperatures (Fig.\ref{fig:2}(d)). At higher fields, for example at $H = 50~\mathrm{kOe}$, the magnetization again increases upon cooling below $T_N$, indicating a non-monotonic field evolution of the low-temperature magnetic response. For magnetic fields above 70 kOe (not shown), the system approaches a polarized paramagnetic state, as evidenced by the monotonic increase of magnetization with decreasing temperature and the absence of a well-defined magnetic transition.

Figure \ref{fig:2}(e) shows $\textit{M}(\textit{H})$ data at $T = 2~\mathrm{K}$ for magnetic fields applied perpendicular (turquoise) and along (purple) $[001]$ direction. For $H \perp [001]$, the magnetization exhibits three distinct metamagnetic transitions, indicating a complex field-driven evolution of the magnetic state. These transitions are further highlighted by pronounced anomalies in the field derivative $\textit{d}(\textit{M})/\textit{dH}$ (red solid line). Notably, the first and third transitions display clear hysteresis between increasing and decreasing field sweeps (see insets), pointing to their first-order character. At $H = 70~\mathrm{kOe}$, the magnetization remains well below the value expected for full saturation. Measurements extended up to 120 kOe (not shown) confirm that the magnetization reaches only about $0.9~\mu_B/Ce$, substantially smaller than the theoretical saturation moment of $2.14~\mu_B/Ce$ for a free Ce$^{3+}$ ion. Such reduced saturation is commonly observed in Ce-based Kondo-lattice systems and can be attributed to a combination of crystal electric field effects and partial Kondo screening. Similar behavior has been reported for other members of the \ce{Ce3\textit{MPn}5} family, including \ce{Ce3TiSb5} \cite{he2025rich} and \ce{Ce3ScBi5} \cite{xu2024magnetization}, as well as in a number of other Ce-based intermetallic compounds, such as \ce{CeMg3} \cite{das2011magnetic}, \ce{CeAgAs2} \cite{mondal2018magnetocrystalline}, \ce{Ce2NiAl6Si5} \cite{zhang2024structural}, and the \ce{Ce\textit{M}Al7Ge4} family \cite{ghimire2016physical}. 

In contrast to the $\textit{M}(\textit{H})$ behavior for $H \perp [001]$, the magnetization measured for $H \parallel [001]$ is significantly smaller and remains nearly linear over the entire investigated field range, reaffirming the $[001]$ direction as the magnetic hard axis. This pronounced magnetic anisotropy primarily reflects strong crystal electric field effects, which confine the Ce moments predominantly within the basal plane. In addition, as discussed previously for \ce{Ce3TiSb5} \cite{he2025rich} and \ce{Ce3TiBi5} \cite{hayami2022magnetic}, the absence of local inversion symmetry at the Ce sites, together with strong spin–orbit coupling, allows for a finite Dzyaloshinskii–Moriya interaction. Such antisymmetric exchange interactions can further influence the in-plane spin configurations and contribute to the anisotropic magnetic response observed in \ce{Ce3MgBi5}.

The temperature evolution of the isothermal magnetization for $H \perp [001]$ is shown in Fig.\ref{fig:2}(f). With increasing temperature, the metamagnetic transitions become progressively weaker and shift toward lower magnetic fields, reflecting the gradual suppression of antiferromagnetic exchange interactions by thermal fluctuations. For temperatures above $T_N$, the magnetization varies nearly linearly with the applied field, characteristic of a paramagnetic response dominated by field-induced polarization of the Ce moments. Taken together, the magnetic measurements in Fig.\ref{fig:2} reveal rich field–temperature phase behavior with multiple field-induced regimes within the magnetically ordered state. In the following, we turn to electrical transport measurements to examine how this magnetic complexity is reflected in the charge transport properties.

The temperature dependence of the electrical resistivity $\rho(T)$ of \ce{Ce3MgBi5}, measured in zero magnetic field, exhibits characteristic features of a Ce-based Kondo-lattice system (Fig.\ref{fig:3}(a)). At high temperatures, $\rho(T)$ exhibits a weak temperature dependence, remaining nearly constant upon cooling from room temperature down to about $150~\mathrm{K}$. Below this temperature, the resistivity develops a broad maximum at approximately $50~\mathrm{K}$, followed by a second, more pronounced maximum near $10~\mathrm{K}$. Such behavior is characteristic of Ce-based Kondo-lattice systems and is commonly attributed to the interplay between Kondo scattering of the localized Ce $4f$ moments and crystal electric field effects. Upon further cooling, $\rho(T)$ decreases sharply, signaling the onset of long-range antiferromagnetic order. The Néel temperature of $T_N = 4.16~\mathrm{K}$, determined from the temperature derivative $\mathrm{d}\rho/\mathrm{d}T$ shown in the inset of Fig.~\ref{fig:3}(a), is in good agreement with the transition temperature obtained from magnetization measurements.

To gain further insight into the low-temperature transport behavior associated with the antiferromagnetic transition, Fig.\ref{fig:3}(b) shows an enlarged view of the resistivity in the vicinity of $T_N$ in zero magnetic field. Below the transition, $\rho(T)$ decreases rapidly, reflecting the reduction of spin-disorder scattering upon the establishment of long-range antiferromagnetic order. The sharp feature at $T_N$ is consistent with the anomaly observed in magnetization and confirms the coupling between magnetic order and charge transport in \ce{Ce3MgBi5}. While an ideal easy-plane antiferromagnet would support a gapless mode, additional in-plane anisotropy, anisotropic exchange, or hybridization effects can lift this degeneracy and open a finite gap. To examine the possibility of such a gap in \ce{Ce3MgBi5}, the low-temperature resistivity was analyzed using a standard model for gapped antiferromagnetic spin-wave scattering \cite{fontes1999electron,zhang2020structural}:

\begin{equation}
\begin{aligned}
\rho(T) =\;& \rho_0 + AT^2 
+ b\,\Delta^2 \sqrt{\frac{T}{\Delta}}
\exp\!\left(-\frac{\Delta}{T}\right)
\times \\
& \left[
1 + \frac{2T}{3\Delta}
+ \frac{2}{15}\left(\frac{T}{\Delta}\right)^2
\right],
\end{aligned}
\label{eq:AFM_spinwave}
\end{equation}

where $\rho_0$ is the residual resistivity, the second term corresponds to the Fermi-liquid contribution, and the third term $\rho_{mag}$) represents electron scattering from antiferromagnetic spin-wave excitations characterized by a finite spin-wave gap $\Delta$. The fit was performed in the antiferromagnetic state for $T \geq 1.9$ K and below $T_{\mathrm{N}}$. Given the limited temperature range, the gap $\Delta$ was fixed to the value obtained independently from the heat-capacity analysis (see below), thereby constraining the fit, reducing parameter uncertainty, and ensuring self-consistency between the heat-capacity and resistivity analyses. Using $\Delta$ = 4.96 K, the fit yields $\rho_0 = 30.5(5)\,\mu\Omega\,\mathrm{cm}$,
$A = 1.13(3)\,\mu\Omega\,\mathrm{cm\,K^{-2}}$, and
$b = 0.69(4)\,\mu\Omega\,\mathrm{cm\,K^{-2}}$. The individual contributions to the resistivity are shown in Fig.\ref{fig:3}(b). To illustrate their relative temperature dependences, the model was extrapolated below the lowest measured temperature, indicating that the $AT^2$ term remains dominant in the low-temperature limit. Consistently, the obtained $A$ coefficient is comparable with values reported for Ce-based Kondo-lattice antiferromagnets, such as \ce{Ce2IrGa12} \cite{zhang2020structural}, reflecting enhanced quasiparticle scattering associated with electronic correlations.

To examine the influence of magnetic field on the transport behavior, Fig.~\ref{fig:3}(c) compares the temperature dependence of the resistivity $\rho(T)$ measured at $H = 0$ and $70~\mathrm{kOe}$, plotted on a logarithmic temperature scale. First, in zero magnetic field $\rho(T)$ exhibits an approximately linear dependence on ln$T$ over an intermediate temperature range above the magnetic ordering temperature and below the broad high-temperature maximum (approximately 10–50 K). This logarithmic behavior has been discussed in the context of Kondo lattices with strong crystal electric field splitting and falls within the phenomenology described by the Cornut–Coqblin framework \cite{cornut1972influence}. Second, the application of a magnetic field strongly modifies the low-temperature transport response: the resistivity maximum near $10~\mathrm{K}$, clearly visible in zero field, is significantly suppressed, indicating a field-induced reduction of magnetic scattering. In addition, the resistive anomaly associated with the antiferromagnetic transition is weakened under applied field, consistent with the suppression of long-range magnetic order observed in the magnetization measurements.

Figure \ref{fig:3}(d) shows the magnetoresistance MR, defined as [$\rho$(\textit{H})-$\rho$(0)]/$\rho$(0), measured at $T = 2~\mathrm{K}$ for $H \perp [001]$. The MR exhibits a nonmonotonic field dependence with three distinct anomalies that closely correspond to the metamagnetic transitions observed in the magnetization $M(H)$ data (see the dashed lines). Changes in slope near the characteristic transition fields indicate that the electronic transport responds sensitively to the field-induced changes of the magnetic state. At higher fields, the MR becomes strongly negative, reaching a value of approximately $-25\%$ at $H=90~\mathrm{kOe}$, which likely results from the suppression of fluctuations of the localized spins by the magnetic field.

The Hall resistivity $\rho_{yx}(H)$ measured at $T=2~\mathrm{K}$ is presented in Fig.\ref{fig:3}(e). The Hall signal displays a clear nonlinear dependence on magnetic field, deviating from the simple linear behavior expected for an ordinary Hall response. In general, the Hall resistivity can be expressed as a sum of ordinary, anomalous, and possible topological contributions \cite{nagaosa2010anomalous,du2021nonlinear,zeng2006linear}. In the present study, a reliable separation of these components is not straightforward due to the complex field evolution of the magnetic structure. Notably, the nonlinearity of $\rho_{yx}(H)$ closely follows the sequence of metamagnetic transitions revealed by the magnetization measurements, pointing to an additional magnetically driven component in the Hall response.

The temperature evolution of the magnetotransport response is summarized in Fig.\ref{fig:3}(f) and Fig.\ref{fig:3}(g), which present the magnetoresistance and Hall resistivity measured at several temperatures around the magnetic ordering region. With increasing temperature, the characteristic anomalies in the MR curves gradually broaden and shift, while their amplitudes are progressively reduced, reflecting the weakening of the field-induced magnetic transitions as thermal fluctuations destabilize the ordered state. A similar temperature evolution is observed in the Hall response (Fig.\ref{fig:3}(g)), where the pronounced nonlinearity of $\rho_{yx}(H)$ at low temperatures progressively diminishes, and the Hall signal approaches an approximately linear field dependence above $T_N$. This trend is further supported by the data at $T = 15$ K, where $\rho_{yx}(H)$ is nearly linear over the measured field range. The nonlinear Hall signal is therefore closely tied to the magnetically ordered and field-induced states and becomes suppressed in the paramagnetic regime.

\begin{figure}[t]
\centering
\includegraphics[width=1\columnwidth]{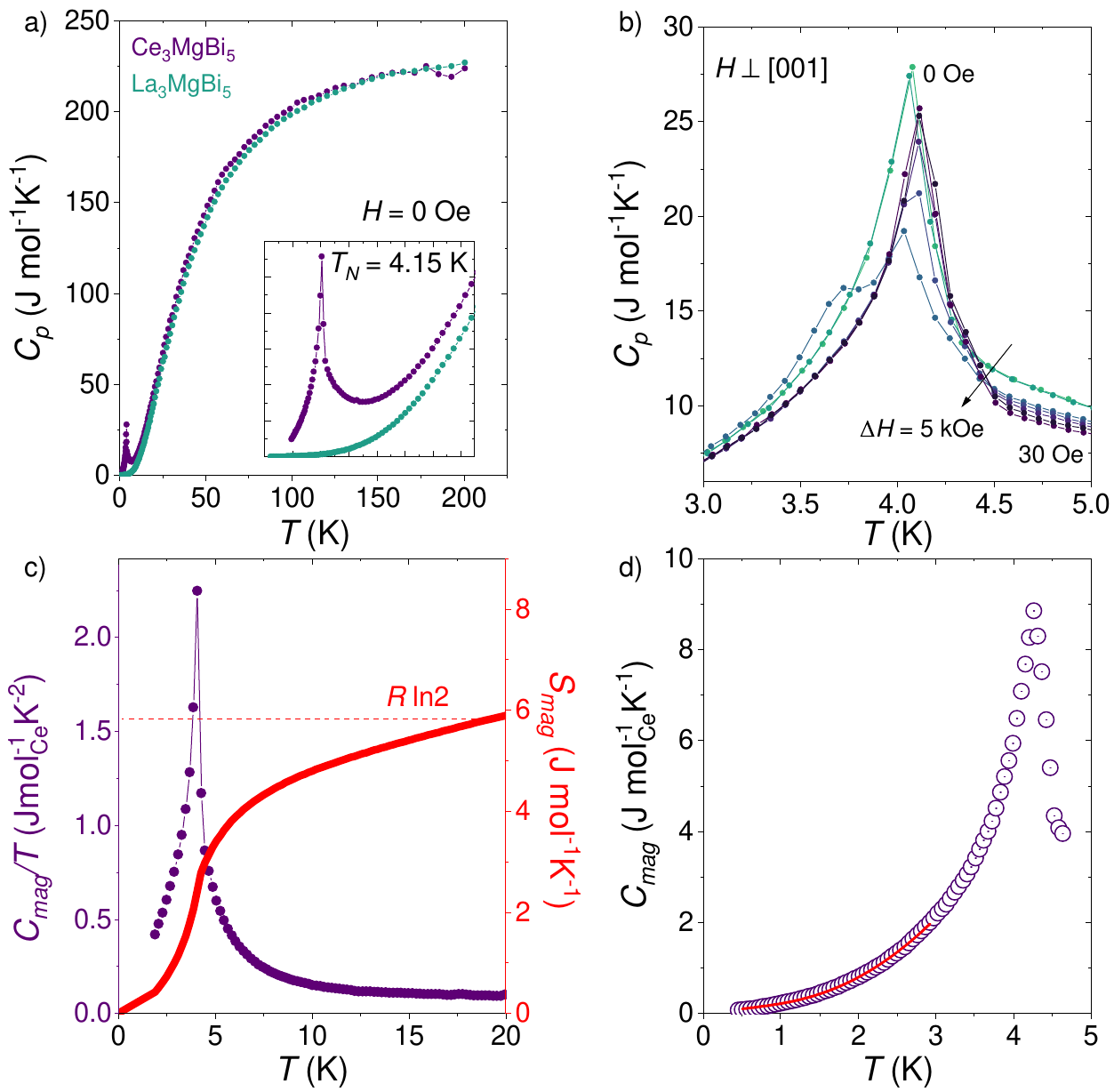}
\caption{(a) Temperature dependence of the specific heat $C_p(T)$ of \ce{Ce3MgBi5} and the nonmagnetic reference compound \ce{La3MgBi5} measured in zero magnetic field over a wide temperature range. The inset highlights the low-temperature region, where a pronounced anomaly at $T_N = 4.15$ K is observed for \ce{Ce3MgBi5}. (b) Low-temperature specific heat of \ce{Ce3MgBi5} measured for $H \perp [001]$ in magnetic fields up to 30 kOe ($\Delta H = 5~\mathrm{kOe}$). (c) Magnetic contribution to the specific heat plotted as $C_{\mathrm{mag}}/T$ together with the corresponding magnetic entropy $S_{\mathrm{mag}}(T)$, obtained by subtracting the phonon background estimated from \ce{La3MgBi5}. The dashed line indicates the value $R\ln2$ expected for a fully localized Kramers doublet. (d) Low-temperature magnetic specific heat $C_{\mathrm{mag}}(T)$ of \ce{Ce3MgBi5}. The solid line represents a fit below $0.65~T_N$ using a model that includes a linear electronic term and a gapped antiferromagnetic spin-wave contribution (see Eq.(2)}).
\label{Fig.4}
\end{figure}

To further characterize the thermodynamic properties of \ce{Ce3MgBi5}, we measured the specific heat $C_p(T)$ in zero magnetic field over a wide temperature range and compared it with the nonmagnetic reference compound \ce{La3MgBi5},  as shown in Fig.\ref{Fig.4}(a). At high temperatures, the specific heat of both compounds approaches the Dulong–Petit limit ($3nR \approx 225~\mathrm{J\,mol^{-1}\,K^{-1}}$, where \textit{n} = 9 is the number of atoms per formula unit, and \textit{R} is the gas constant), confirming the expected stoichiometry. Upon cooling, a pronounced anomaly appears in $C_p(T)$ of \ce{Ce3MgBi5} at $T_N \approx 4.15~\mathrm{K}$, signaling the onset of long-range antiferromagnetic order. This feature, highlighted in the inset of Fig.\ref{Fig.4}(a), is absent in \ce{La3MgBi5}, consistent with its nonmagnetic ground state. The transition temperature extracted from the specific heat is in good agreement with the Néel temperature determined from magnetization and transport measurements. The response of the antiferromagnetic transition to an applied magnetic field provides further insight into the stability and field evolution of the ordered state. As shown in Fig.\ref{Fig.4}(b), the low-temperature specific heat of \ce{Ce3MgBi5} measured for $H \perp [001]$ exhibits a systematic broadening of the anomaly with increasing field. The position of the main transition shows only a weak field dependence and is shifted slightly toward higher temperatures, rather than being strongly suppressed. In addition, for fields of the order of $H\sim10~\mathrm{kOe}$, a second anomaly becomes visible in $C_p(T)$, indicating the emergence of a field-induced magnetic transition within the ordered phase. This behavior is consistent with the multiple field-driven transitions inferred from magnetization and magnetotransport measurements and points to a complex evolution of the antiferromagnetic state under the applied magnetic field.

To extract the magnetic contribution to the specific heat, the phonon background was estimated using the isostructural \ce{La3MgBi5} and subtracted from the total specific heat of \ce{Ce3MgBi5}. The resulting magnetic specific heat, plotted as $C_{\mathrm{mag}}/T$ in Fig.\ref{Fig.4}(c), exhibits a clear anomaly at $T_N$, confirming the bulk nature of the antiferromagnetic transition. The magnetic entropy was obtained by integrating $C_{\mathrm{mag}}/T$ over temperature, $S_{\mathrm{mag}}(T) = \int_0^T \frac{C_{\mathrm{mag}}(T)}{T} \, \mathrm{d}T$, with the low-temperature contribution estimated by smoothly extrapolating toward $T \rightarrow 0$. As shown in Fig.\ref{Fig.4}(c), the full $R\ln2$ entropy associated with a Kramers doublet is recovered by 20 K, indicating an extended temperature range of magnetic fluctuations consistent with Kondo correlations affecting the Ce $4f$ moments. Similar reduced magnetic entropy has been reported in several Ce-based Kondo-lattice antiferromagnets \cite{zhang2024structural,xie2024complex, ghimire2016physical}, including related members of the \ce{Ce3\textit{MPn}5} family \cite{fu2024high, xu2024magnetization}, highlighting the importance of hybridization effects for the low-temperature magnetic properties of these systems. 

Figure \ref{Fig.4}(d) shows the magnetic specific heat of \ce{Ce3MgBi5}, $C_{\mathrm{mag}}(T)$, in the vicinity of $T_N$. The data below approximately $0.65~T_N$ were fitted using a model that includes a linear electronic contribution, $\gamma T$, and a gapped antiferromagnetic spin-wave term:

\begin{equation}
\begin{aligned}
C_{\mathrm{mag}}(T) = \;& \gamma T 
+ c\,\Delta^{7/2}\sqrt{T}\,
\exp\!\left(-\frac{\Delta}{T}\right)
\times \\
& \left[
1 + \frac{39}{20}\left(\frac{T}{\Delta}\right)
+ \frac{51}{32}\left(\frac{T}{\Delta}\right)^2
\right],
\end{aligned}
\end{equation}

where $\gamma$ is the electronic Sommerfeld coefficient, $c$ represents a prefactor related to the spin-wave stiffness, and $\Delta$ is the energy gap in the antiferromagnetic spin-wave spectrum. The fit gives a Sommerfeld coefficient $\gamma = 199(5)~\mathrm{mJ~mol^{-1}_{Ce}K^{-2}}$ and a spin-wave gap $\Delta = 4.96(19)$~K. The enhanced value of $\gamma$ points to a substantial low-energy electronic density of states, consistent with Kondo hybridization persisting into the magnetically ordered phase. A similarly enhanced Sommerfeld coefficient has been reported for the related compound \ce{Ce3TiBi5} ($\gamma \approx 210~\mathrm{mJ~mol^{-1}_{Ce}K^{-2}}$) \cite{motoyama2018magnetic}, placing \ce{Ce3MgBi5} in a comparable regime of electronic correlations. For completeness, we also estimated $\gamma$ from a linear fit $C/T$ vs $T^2$ in a narrow temperature range above $T_N$ (not shown). Depending on the fitting window, this analysis yields $\gamma$ values in the range of 200-225$~\mathrm{mJ~mol^{-1}_{Ce}K^{-2}}$, consistent with the value obtained from the low-temperature $C_{mag}$ fit. Within the Fermi-liquid framework, the electronic specific heat and resistivity are governed by quasiparticles with a renormalized effective mass, $\gamma \propto m^*$ and $A \propto (m^*)^2$, leading to an approximately constant Kadowaki–Woods ratio $R_{\mathrm{KW}} = A/\gamma^2 = 1.0 \times 10^{-5}\ \mu\Omega\,\mathrm{cm}\,\mathrm{mol}^2\,\mathrm{K}^2\,\mathrm{mJ}^{-2}$ for a broad class of strongly correlated materials \cite{kadowaki1986universal, jacko2009unified}. Using $\gamma$ from heat-capacity data and the $A$ coefficient from resistivity analysis, we obtain $A/\gamma^2 \approx 2.9 \times 10^{-5}$ $\mu\Omega\,\mathrm{cm}\,\mathrm{mol}^2\,\mathrm{K}^2\,\mathrm{mJ}^{-2}$, which falls within the expected range for correlated Ce-based systems \cite{kadowaki1986universal, falkowski2015cooperative}. 

The extracted spin-wave gap indicates a finite excitation gap in the antiferromagnetic state. Recent neutron-scattering studies on the related compounds \ce{Ce3TiBi5} and \ce{Ce3ZrBi5} revealed well-separated CEF excitations, demonstrating that the \ce{Ce^{3+}} ground state in this family is a well-isolated Kramers doublet corresponding to an effective pseudospin-1/2 degree of freedom \cite{park2026magneticorderexcitationsce3tibi5}. Within such a scenario, the presence of a finite spin-wave gap would not be expected to originate from a conventional single-ion anisotropy term and may instead reflect anisotropic exchange interactions or hybridization effects influencing the low-energy spin dynamics of \ce{Ce3MgBi5}, although further microscopic probes will be required to establish the precise origin of this gap.

\begin{figure}[t]
\centering
\includegraphics[width=0.95\columnwidth]{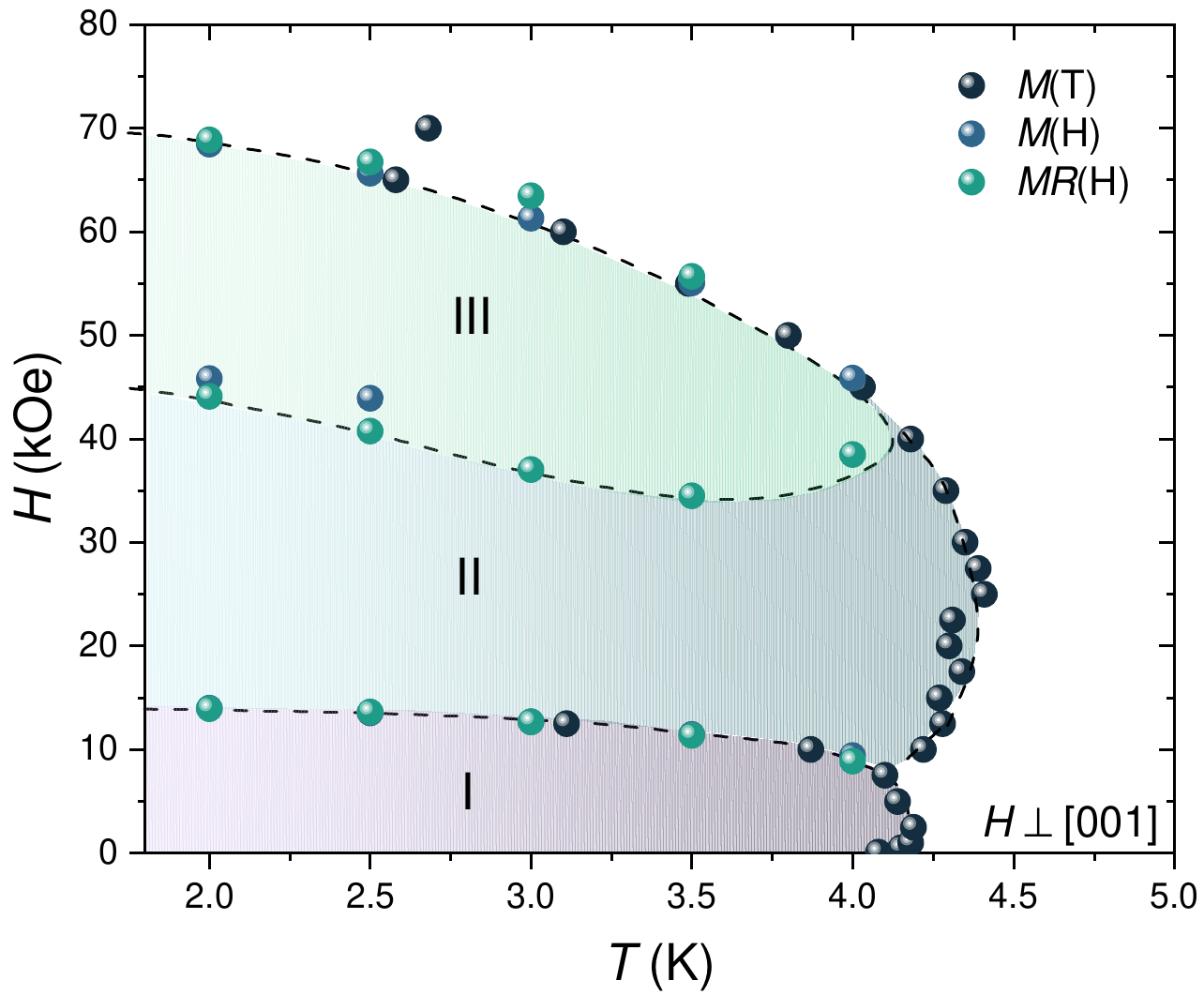}
\caption{Magnetic field–temperature phase diagram of \ce{Ce3MgBi5} for magnetic fields applied perpendicular to the $[001]$ direction. Dashed lines are guides to the eye.}
\label{Fig.5}
\end{figure}

Based on anomalies identified in the temperature-dependent magnetization $M(T)$, isothermal magnetization curves $M(H)$, and magnetotransport measurements, a magnetic field–temperature ($H-T$) phase diagram for \ce{Ce3MgBi5} was constructed, as shown in Fig.\ref{Fig.5}. Because the magnetization for magnetic fields applied along the $[001]$ direction remains small and nearly linear over the entire investigated field range - confirming $[001]$ as the magnetic hard axis - we focus here on the phase diagram for $H \perp [001]$, where the magnetic response is most pronounced. 

In zero magnetic field, \ce{Ce3MgBi5} undergoes a transition into an antiferromagnetically ordered state at $T_N \approx 4.2~\mathrm{K}$. Upon application of a magnetic field perpendicular to the $[001]$ direction, the ordering temperature shows a non-monotonic evolution and defines a dome-like ordered region in the $H-T$ plane, rather than a simple field-driven suppression of antiferromagnetism. This behavior reflects a competition between magnetic exchange interactions and the Zeeman energy in a strongly anisotropic system.

Within the ordered region, additional field-induced anomalies subdivide the low-temperature phase space into three distinct regimes, labeled I–III in Fig.\ref{Fig.5}. The boundaries between these regimes are defined by critical fields that are consistently observed as metamagnetic steps in the $M(H)$ curves and as corresponding anomalies in the magnetoresistance. The presence of multiple field-driven transitions within the antiferromagnetically ordered state indicates a nontrivial evolution of the magnetic ground state under the applied field and suggests that several magnetic configurations with comparable energies are stabilized over different field ranges rather than a simple antiferromagnetic structure. Such multi-stage magnetic phase diagrams have also been reported for other members of the \ce{Ce3\textit{M}\textit{Pn}5} family, including \ce{Ce3TiSb5}, \ce{Ce3TiBi5}, and \ce{Ce3ScBi5}. In these systems, the combination of geometric frustration associated with the quasi-kagome arrangement of Ce moments, anisotropic exchange interactions, and strong crystal electric field effects can give rise to multiple field-induced magnetic transitions for in-plane fields. In \ce{Ce3MgBi5}, the reduced saturation moment and signatures of Kondo hybridization observed in transport and thermodynamic measurements suggest that the field-induced phases involve partially hybridized Ce $4f$ moments. Within this picture, the applied magnetic field likely drives the system through a sequence of magnetically distinct regimes that remain embedded in a Kondo-lattice background, rather than between purely localized-moment configurations.

\section{Summary}
In this study, we successfully synthesized high-quality single crystals of \ce{Ce3MgBi5} and carried out a comprehensive investigation of its structural, magnetic, thermodynamic, and transport properties. \ce{Ce3MgBi5} crystallizes in the hexagonal \textit{P}6$_3$/\textit{mcm} structure, in which the anisotropic arrangement of Ce ions combines zig-zag chains along the crystallographic \textit{c} axis with a distorted kagome-like network in the basal plane, providing a natural setting for competing magnetic interactions.
Magnetization measurements reveal that \ce{Ce3MgBi5} undergoes antiferromagnetic ordering below $T_N \approx 4.2$~K with pronounced magnetic anisotropy, where the $[001]$ direction acts as a hard magnetic axis. For magnetic fields applied within the basal plane, the ordered state evolves in a nontrivial manner, exhibiting multiple metamagnetic transitions and a dome-like evolution of $T_N$ in the $H–T$ phase diagram. The origin of this rich phase behavior remains an open question and likely reflects the combined influence of anisotropic exchange interactions, crystal electric field effects, and Kondo hybridization. Electrical transport measurements display characteristic signatures of a Ce-based Kondo-lattice, including weak temperature dependence at high temperatures and broad resistivity maxima associated with Kondo scattering and crystal electric field effects, followed by a pronounced decrease below $T_N$, signaling the establishment of long-range antiferromagnetic order. Magnetoresistance and Hall effect data reveal clear field-induced anomalies that track the metamagnetic transitions, demonstrating a strong coupling between charge transport and the magnetic state. Specific-heat measurements confirm the magnetic transition and show that the magnetic entropy is gradually released over a broad temperature range above $T_N$, consistent with persistent Kondo correlations. 

To summarize, our results establish \ce{Ce3MgBi5} as a new member of the \ce{Ce3\textit{M}\textit{Pn}5} family, in which antiferromagnetic order may emerge from the interplay of geometric frustration, strong magnetic anisotropy, and Kondo-lattice effects. The combination of multiple field-induced magnetic transitions, reduced magnetic entropy, and correlated transport responses underscores the complexity of its magnetic ground state. Further experimental and theoretical studies will be required to fully elucidate the microscopic magnetic structure and the role of competing interactions in this system. 

\section{Acknowledgment}
This work was supported by the U.S. Department of Energy, Office of Science, Basic Energy Sciences, Materials Sciences and Engineering Division. 

\section{Data Availability}
The data that support the findings of this article are openly available \cite{Data}.

\end{document}